\documentclass[12pt,a4paper]{article}
\usepackage[utf8]{inputenc}
\usepackage[english]{babel}
\usepackage{amsmath}
\usepackage{amsfonts}
\usepackage{amssymb}
\usepackage[margin=1in]{geometry}
\usepackage{amsthm}
\usepackage{hyperref}
\usepackage[numbers]{natbib}

\newtheorem{df}{Definition}[section]
\newtheorem{lem}{Lemma}[section]
\newtheorem{prp}{Proposition}[section]
\newtheorem{thm}{Theorem}[section]

\numberwithin{equation}{section}

\title{The Hull-White Model under Volatility Uncertainty}
\author{Julian H\"olzermann\footnote{Center for Mathematical Economics, Bielefeld University. Email: julian.hoelzermann@uni-bielefeld.de. The author thanks Frank Riedel for valuable advice and Tolulope Fadina, Damir Filipovi\'c, Jan Ob\l\'oj, and Thorsten Schmidt for fruitful conversations. The author gratefully acknowledges financial support by the German Research Foundation (Deutsche Forschungsgemeinschaft) via Collaborative Research Center 1283.}}

\begin{document}

\maketitle

\begin{abstract}
\noindent We study the Hull-White model for the term structure of interest rates in the presence of volatility uncertainty. The uncertainty about the volatility is represented by a set of beliefs, which naturally leads to a sublinear expectation and a $G$-Brownian motion. The main question in this setting is how to find an arbitrage-free term structure. This question is crucial, since we can show that the classical approach, martingale modeling, does not work in the presence of volatility uncertainty. Therefore, we need to adjust the model in order to find an arbitrage-free term structure. The resulting term structure is affine with respect to the short rate and the adjustment factor. Although the adjustment changes the structure of the model, it is still consistent with the traditional Hull-White model after fitting the yield curve. In addition, we extend the model and the results to a multifactor version, driven by multiple risk factors.
\end{abstract}


\section{Introduction}
Traditional models in finance are subject to model uncertainty and, especially, volatility uncertainty. These models typically assume that there is a single, known probability measure determining the behavior of the fundamental quantities on the market. This assumption simplifies the modeling of financial markets and the pricing of derivatives, but in many situations it is not possible to specify the underlying probability measure of the model. The uncertainty about using the correct probabilistic law is called \textit{model uncertainty}. There is an increasing literature, commonly referred to as \textit{robust finance}, dealing with model uncertainty by investigating financial markets in the presence of a family of possible probability measures or none at all. The overall aim is to make models robust with respect to misspecifications regarding the probabilistic law. An important kind of model uncertainty is volatility uncertainty. The future evolution of the volatility of an underlying is impossible to predict and it is hard to specify its probabilistic law.
\par In the present paper, we study the Hull-White model for the term structure of interest rates under volatility uncertainty. That means, we model the instantaneous spot interest rate, called short rate, in the presence of a family of possible beliefs about the volatility and it is completely uncertain which one is correct. We do not impose any probabilistic assumptions on the beliefs about the volatility, that is, there is no assumption on how likely it is for each belief to be the correct one. We consider all beliefs in which the volatility is bounded by two extreme values. This assumption might be restrictive, but it is sufficient to make the mathematical analysis work. In general, one can relax this assumption with further technical effort and, from an economic perspective, the succeeding results do not require this assumption, as the results do not alter if the bounds for the volatility do.
\par Mathematically speaking, we model the short rate as an Ornstein-Uhlenbeck process driven by a $G$-Brownian motion. As in the classical Hull-White model, we describe the evolution of the short rate by a diffusion process of Ornstein-Uhlenbeck type. The difference compared to the traditional model is that the volatility is uncertain. We represent the uncertainty about the volatility by a set of probability measures, called the \textit{set of beliefs}, where each measure corresponds to a different belief about the volatility. That means, the short rate has a different volatility under each measure and it is uncertain which measure is the correct belief about the volatility. Such a set of measures naturally leads to a sublinear expectation and a $G$-Brownian motion. A sublinear expectation can be interpreted as a worst-case measure. A $G$-Brownian motion is basically a Brownian motion with an uncertain volatility. Thus, we use the $G$-Brownian motion to describe the evolution of the short rate, i.e., as the driver of the short rate dynamics. Due to the uncertainty about its volatility, the short rate has an uncertain variance.
\par The main question in this paper is how to find an arbitrage-free term structure in the presence of volatility uncertainty. The crucial characteristic of volatility uncertainty is that it is represented by a nondominated set of beliefs. That means, there is no measure dominating all measures in the set of beliefs. Hence, it is not possible to find a single equivalent martingale measure for the related bond market. Thus, the discussion about arbitrage becomes a subtle issue in this framework. If we want to follow a martingale modeling approach, we need to choose the bond prices in such a way that the discounted bonds are symmetric $G$-martingales, which means that they are martingales in each possible scenario for the volatility. Martingale modeling is a common approach in short rate models, but we can unfortunately show that this approach does not work under the initially given set of beliefs.
\par In order to find an arbitrage-free term structure, we consider sublinear expectations defined by a linear $G$-backward stochastic differential equation. By standard results on $G$-backward stochastic differential equations, we can define consistent sublinear expectations by this procedure. Since the $G$-backward stochastic differential equation is linear, there exists an explicit solution. The representation of the solution shows that the resulting sublinear expectation corresponds to the expectation under an equivalent change of measure. We can also formally show that sublinear expectations defined in this way are in some sense equivalent to the initial one. As a consequence, the bond market is arbitrage-free if there exists a sublinear expectation of this particular type under which the discounted bonds are symmetric $G$-martingales.
\par We show that there exists a sublinear expectation of the above kind under which there is a unique arbitrage-free term structure. If we choose a particular process as a coefficient in the linear $G$-backward stochastic differential equation defining equivalent sublinear expectations, we obtain a sublinear expectation under which there is a unique expression for the bond prices such that the discounted bonds are symmetric $G$-martingales. The choice might seem special, but it can be justified by economic arguments. Due to the Girsanov transformation for $G$-Brownian motion, the process represents an adjustment factor, adjusting the short rate by its uncertain variance. Alternatively, we can interpret the process as the market price of risk. Since the model is not only subject to risk but also subject to uncertainty, we also refer to the process as the \textit{market price of uncertainty}. The resulting bond prices are different, though similar, to the prices from the traditional model without volatility uncertainty. In particular, they have an affine structure with respect to the short rate and the market price of uncertainty.
\par Even though the structure of the model is different from the traditional one, we are yet consistent with the classical Hull-White model after fitting the yield curve. Since we consider an equivalent sublinear expectation, the Girsanov transformation for $G$-Brownian motion implies that the dynamics of the short rate, as well as the bond prices, differ from the ones of the traditional model. As in the classical model, we use the mean reversion level of the short rate to fit the bond prices of the model to an initial yield curve, observable on the market. Surprisingly, then the short rate dynamics and the bond prices are again consistent with the ones from the classical Hull-White model. They are consistent in the sense that the short rate dynamics and the bond prices coincide with the classical ones if we drop the uncertainty about the volatility.
\par In addition, we study an extension of the model driven by multiple risk factors. For the sake of simplicity, we derive the results mentioned above in the presence of a single risk factor, that is, the short rate is driven by a single $G$-Brownian motion. Such a framework simplifies the interpretation and the intuition of the results and enables us to compare the results with the classical Hull-White model. However, empirical studies show that more factors are needed in order to explain term structure movements \citep*{adriancrumpmoench2013,daisingleton2003,joslinpriebschsingleton2014}. Therefore, we consider a model extension in which the short rate is affected by several risk factors with uncertain volatilities and uncertain correlations. We are able to extend all of the previous results to the general case.
\par Regarding the literature, the concept of model uncertainty originates from \textit{Knightian uncertainty}, which was introduced by \citet*{knight1921}. Knightian uncertainty is considered to be the counterpart of risk, because risk can be measured by a probability in contrast to uncertainty. Such uncertainty describes events in reality which are too complex or for which the related data are missing to calculate the risk of such an event. In such situations, one usually assumes that some quantities are completely uncertain but bounded by two extreme values. In the case of volatility uncertainty, we accordingly consider all volatility scenarios which take values in a certain interval without any further specification.
\par There are many approaches to describe volatility uncertainty mathematically. On the one hand, there are the approaches of \citet*{denismartini2006} and \citet*{sonertouzizhang2011b}, which start from a probabilistic point of view. On the other hand, there is the theory of $G$-Brownian motion from \citet*{peng2019}, in which distributions are characterized by nonlinear partial differential equations. Indeed, the approaches of \citet*{denismartini2006} and \citet*{peng2019} are connected by a duality, which was shown by \citet*{denishupeng2011}. Moreover, there are several generalizations, such as an extension to time dependent, stochastic bounds for the family of volatility scenarios \citep*{nutz2013} and a pathwise stochastic calculus \citep*[and references therein]{contperkowski2019}. We mainly use the theory of $G$-Brownian motion in this paper, since the literature on $G$-Brownian motion offers a lot of results.
\par The literature on model uncertainty in asset markets is very extensive. There are many works on volatility uncertainty in asset markets \citep*{avellanedalevyparas1995,epsteinji2013,lyons1995,vorbrink2014}. Since volatility uncertainty is represented by a nondominated set of probability measures, the classical fundamental theorem of asset pricing cannot be used in such a situation. Therefore, there are many approaches to extend the theorem to a multiprior setting \citep*{bayraktarzhou2017,biaginibouchardkardarasnutz2017,bouchardnutz2015}. In addition, there are also model-free versions, i.e., without any reference measure \citep*{acciaiobeiglbockpenknerschachermayer2016,burzonifrittellihoumaggisobloj2019,riedel2015}. Another extensively studied topic related to robust finance is the pricing and hedging of contingent claims in the presence of multiple priors \citep*[and references therein]{aksamitdengoblojtan2019,carassusoblojwiesel2019,possamairoyertouzi2013} or without any priors at all \citep*[and references therein]{bartlkupperpromeltangpi2019,beiglbockcoxhuesmannperkowskipromel2017,schiedvoloshchenko2016}.
\par The literature on model uncertainty in interest rate models is relatively sparse and there is no paper focusing on obtaining an arbitrage-free term structure in the presence of volatility uncertainty. \citet*{epsteinwilmott1999} study an interest rate model with no underlying probabilistic assumptions. Instead, they use Knightian uncertainty to describe the evolution of the short rate, which typically leads to a range of prices for contracts on the interest rate. Unfortunately, the discussion about the absence of arbitrage is, like all of the derivations, very intuitive and less mathematical. \citet*{avellanedalewicki1996} adapt the principle of volatility uncertainty to construct an interest rate model that also leads to a range of prices. However, the absence of arbitrage is treated in a very intuitive way as well. \citet*{fadinaschmidt2019} introduce a term structure model with default risk and ambiguity. The ambiguous parameter is the default intensity and the set of priors is dominated. This simplifies the discussion about arbitrage and results in a range of no-arbitrage prices for defaultable bonds. \citet*{fadinaneufeldschmidt2019} study affine processes with parameter uncertainty and the pricing of bonds under ambiguity, which also results in a range of prices. However, this procedure is justified by a superhedging argument from \citet*{biaginibouchardkardarasnutz2017}, which is not reasonable for pricing bonds, since they are the fundamental quantities in fixed income markets, i.e., they cannot be hedged.
\par A different, though related, approach to overcome the stylized facts about the volatility of the short rate is to use regime switching models. Regime switching term structure models assume that the volatility follows a continuous-time Markov chain, which jumps between a finite number of values. The literature on regime switching term structure models shows that these models offer advantages compared to classical term structure models \citep*{daisingletonyang2007,gourierouxmonfortpegorarorenne2014,monfortpegoraro2007}. In the present paper, we consider all volatility processes bounded by two extreme values. Thus, we also consider trajectories described by regime switching term structure models. However, this is just a pathwise argument. One needs to enlarge the probability space of the model in order to obtain the same Markov chain considered by regime switching term structure models, since the probability of the Markov chain jumping to a different state is usually independent from the remaining risk factors.
\par In general, volatility uncertainty is a conceptually different approach compared to stochastic volatility. There is plenty of evidence that the volatility of all major financial quantities is neither constant nor deterministic. This problem is addressed by stochastic volatility models, in which the volatility of an underlying is modeled by a particular stochastic process. The latter is typically chosen such that it shares the characteristics of the historical volatility and the model-implied option prices match the prices observed on the market. However, there could be many model specifications performing this task. Moreover, it is not sure if a volatility specification which is consistent with the past is still valid in the future, since the market environment can change drastically. The goal of considering an uncertain volatility is to loosen the assumption that the probabilistic law of the volatility is perfectly known by considering a family of possible probabilistic laws without knowing which one is correct in order to make the model robust.
\par The paper is organized as follows. In Section \ref{short rate dynamics}, we construct a framework for modeling volatility uncertainty and introduce the short rate process. Section \ref{related bond market} introduces the related bond market. In Section \ref{martingale modeling}, we adapt the concept of martingale modeling to volatility uncertainty and show that martingale modeling does not work in the presence of volatility uncertainty. Hence, we define equivalent sublinear expectations in Section \ref{equivalent sublinear expectations}, which we can use to find an arbitrage-free term structure. In Section \ref{arbitrage-free term structure}, we show that there exists an equivalent sublinear expectation under which we obtain a unique arbitrage-free term structure. Section \ref{yield curve fitting} shows how to fit the model to an initially observed term structure. In Section \ref{multifactor extension}, we extend the model to a version driven by multiple risk factors. Section \ref{conclusion} gives a conclusion.

\section{Short Rate Dynamics}\label{short rate dynamics}
In the traditional Hull-White model, without volatility uncertainty, the short rate is described by an Ornstein-Uhlenbeck process driven by a standard Brownian motion. Let us consider a probability space $(\Omega,\mathcal{F},P_0)$ such that $\Omega=C_0(\mathbb{R}_+)$, $\mathcal{F}=\mathcal{B}(\Omega)$, and $P_0$ is the Wiener measure. $C_0(\mathbb{R}_+)$ and $\mathcal{B}(\Omega)$ denote the space of all $\mathbb{R}$-valued continuous paths on $\mathbb{R}_+$ starting in $0$ and the Borel $\sigma$-algebra on $\Omega$, respectively. We denote by $B=(B_t)_{t\geq0}$ the canonical process and we choose $\mathbb{F}=(\mathcal{F}_t)_{t\geq0}$ to be the filtration generated by $B$ completed by all $P$-null sets. Then $B$ is a standard Brownian motion under $P_0$. The classical Hull-White model, without volatility uncertainty, assumes that the short rate process $r=(r_t)_{t\geq0}$ satisfies the stochastic differential equation
\begin{align}\label{short rate dynamics without volatility uncertainty}
r_t=r_0+\int_0^t\big(\mu(u)-\alpha r_u\big)du+\sigma B_t
\end{align}
for a suitably integrable function $\mu:\mathbb{R}_+\rightarrow\mathbb{R}$ and constants $\alpha,\sigma>0$. So the short rate is a mean reverting process with a constant volatility.
\par In the presence of volatility uncertainty, we consider a family of probability measures, where each measure represents a different belief about the volatility. The state space for the volatility is given by the interval $[\underline{\sigma},\overline{\sigma}]$ for $\overline{\sigma}\geq\underline{\sigma}>0$. We denote by $\mathcal{A}$ the collection of all $[\underline{\sigma},\overline{\sigma}]$-valued, $\mathbb{F}$-adapted processes. That is, $\mathcal{A}$ consists of all possible volatility processes bounded by two extreme values. For each $\sigma=(\sigma_t)_{t\geq0}\in\mathcal{A}$, we define the process $B^\sigma=(B_t^\sigma)_{t\geq0}$ by,
\begin{align*}
B_t^\sigma:=\int_0^t\sigma_udB_u,
\end{align*}
and we define the measure $P^\sigma$ to be the law of the process $B^\sigma$, i.e.,
\begin{align*}
P^\sigma:=P_0\circ(B^\sigma)^{-1}.
\end{align*}
We call the collection of all such measures $\mathcal{P}:=\{P^\sigma\vert\sigma\in\mathcal{A}\}$ the \textit{set of beliefs}, since $\mathcal{P}$ contains all beliefs about the volatility. Now the canonical process has a different volatility under each measure in the set of beliefs. In addition, we define the sublinear expectation $\hat{\mathbb{E}}$ as the upper expectation of the set of beliefs,
\begin{align*}
\hat{\mathbb{E}}[\xi]:=\sup_{P\in\mathcal{P}}\mathbb{E}_P[\xi]
\end{align*}
for all measurable $\xi$ such that $\mathbb{E}_P[\xi]$ exists for all $P\in\mathcal{P}$. If $\xi$ represents a financial loss, then $\hat{\mathbb{E}}$ can be understood as a risk measure. In general, $\mathbb{E}$ represents a worst-case measure.
\par Volatility uncertainty naturally leads to a $G$-Brownian motion. According to \citet*{denishupeng2011}, $\hat{\mathbb{E}}$ corresponds to the $G$-expectation on $L_G^1(\Omega)$ and the canonical process $B$ is a $G$-Brownian motion under $\hat{\mathbb{E}}$. The $G$-expectation is defined via a nonlinear heat equation, where the nonlinear generator $G:\mathbb{R}\rightarrow\mathbb{R}$ is given by
\begin{align*}
G(a)=\tfrac{1}{2}\sup_{\sigma\in[\underline{\sigma}^2,\overline{\sigma}^2]}\{\sigma a\}.
\end{align*}
$L_G^1(\Omega)$ is the space of random variables for which the $G$-expectation is defined. There are also attempts to enlarge the domain of the $G$-expectation. Here we use the classical space in order to be able to apply all of the results from the literature on $G$-Brownian motion. We identify random variables $\xi,\xi'\in L_G^1(\Omega)$ if they are equal \textit{quasi-surely}, i.e., $P$-almost surely for all $P\in\mathcal{P}$, which is equivalent to $\hat{\mathbb{E}}[\vert\xi-\xi'\vert]=0$. We also use the terminology $\mathcal{P}$\textit{-quasi-surely} if we need to indicate under which set of measures a statement holds quasi-surely. The details regarding the calculus of $G$-Brownian motion can be found in the book of \citet*{peng2019}.
\par A $G$-Brownian motion is tailor-made for modeling volatility uncertainty, since it has no mean uncertainty but variance uncertainty. That means, for all $t$,
\begin{gather*}
\hat{\mathbb{E}}[B_t]=0=-\hat{\mathbb{E}}[-B_t],
\\\hat{\mathbb{E}}[B_t^2]=\overline{\sigma}^2t\geq\underline{\sigma}^2t=-\hat{\mathbb{E}}[-B_t^2].
\end{gather*}
As a consequence, the quadratic variation of a $G$-Brownian motion is an uncertain process, which, for all $t$, satisfies
\begin{align*}
\overline{\sigma}^2t\geq\langle B\rangle_t\geq\underline{\sigma}^2t.
\end{align*}
Thus, the canonical process is unambiguous in mean and evolves with a volatility being at most $\overline{\sigma}$ and at least $\underline{\sigma}$. The previous inequality shows that $B$ is a standard Brownian motion with a constant volatility if there is no volatility uncertainty, that is, if $\overline{\sigma}=\underline{\sigma}$.
\par We describe the behavior of the short rate by an Ornstein-Uhlenbeck process driven by a $G$-Brownian motion. We choose the same structure as in the classical Hull-White model. The difference is that we include volatility uncertainty by replacing the constant volatility and the standard Brownian motion by a $G$-Brownian motion. Hence, the short rate process $r$ is supposed to be given by the $G$-stochastic differential equation
\begin{align}\label{short rate dynamics with volatility uncertainty}
r_t=r_0+\int_0^t\big(\mu(u)-\alpha r_u\big)du+B_t
\end{align}
for a suitably integrable function $\mu:\mathbb{R}_+\rightarrow\mathbb{R}$ and a constant $\alpha>0$. Thus, the short rate has a time dependent mean reversion level, which is deterministic, and a time dependent volatility, which is uncertain. This is desirable, since we can use the mean reversion level for yield curve fitting and we do not have to specify any volatility structure. It should be noted that \eqref{short rate dynamics with volatility uncertainty} corresponds to \eqref{short rate dynamics without volatility uncertainty}, i.e., the classical case without volatility uncertainty, if $\overline{\sigma}=\sigma=\underline{\sigma}$.
\par The $G$-stochastic differential equation describing the dynamics of the short rate has a closed-form solution. By Theorem 5.1.3 from \citet*{peng2019}, we know that \eqref{short rate dynamics with volatility uncertainty} has a unique solution in $M_G^2(0,T)$ for every $T<\infty$. The space $M_G^2(0,T)$ consists of stochastic processes with a certain regularity and is used to define stochastic integrals in the calculus of $G$-Brownian motion. As in the classical case, we can explicitly solve \eqref{short rate dynamics with volatility uncertainty}.
\begin{prp}\label{solution to short rate dynamics}
The solution to the $G$-stochastic differential equation \eqref{short rate dynamics with volatility uncertainty} is given by
\begin{align}\label{representation of the short rate}
r_t=e^{-\alpha t}r_0+\int_0^te^{-\alpha(t-u)}\mu(u)du+\int_0^te^{-\alpha(t-u)}dB_u.
\end{align}
\end{prp}
\begin{proof}
This can be verified by using It\^o's formula for $G$-Brownian motion \citep*[Theorem 5.4]{lipeng2011}. The verification works totally analogous to the classical case with a standard Brownian motion.
\end{proof}
\par The short rate has no mean uncertainty, but it has an uncertain variance. We can easily show that the upper expectation of the short rate coincides with its lower expectation. Thus, the mean of the short rate is deterministic. In addition, we can show that the upper, respectively lower, expectation of the squared deviation of the short rate from its mean is given by the variance from the classical Hull-White model with the highest, respectively lowest, possible volatility. Hence, the short rate has an uncertain variance, which is bounded by two extreme values.
\begin{thm}\label{mean and variance of the short rate}
For all $t$, the short rate $r_t$ satisfies
\begin{subequations}
\begin{gather}
\hat{\mathbb{E}}[r_t]=e^{-\alpha t}r_0+\int_0^te^{-\alpha(t-u)}\mu(u)du=-\hat{\mathbb{E}}[-r_t],\label{mean of the short rate}
\\\hat{\mathbb{E}}\big[(r_t-\hat{\mathbb{E}}[r_t])^2\big]=\tfrac{\overline{\sigma}^2}{2\alpha}(1-e^{-2\alpha t})\geq\tfrac{\underline{\sigma}^2}{2\alpha}(1-e^{-2\alpha t})=-\hat{\mathbb{E}}\big[-(r_t-\hat{\mathbb{E}}[r_t])^2\big].\label{variance of the short rate}
\end{gather}
\end{subequations}
\end{thm}
\begin{proof}
First, we sketch how to obtain \eqref{mean of the short rate}. The first two summands on the right-hand side of \eqref{representation of the short rate} are deterministic. We know that the upper expectation and the lower expectation of an integral with respect to a $G$-Brownian motion vanish. Thus, it holds \eqref{mean of the short rate}.
\par In order to show \eqref{variance of the short rate}, we use the nonlinear Feynman-Kac formula from \citet*{hujipengsong2014}. We define the process $X=(X_t)_{t\geq0}$ as the deviation of $r$ from its mean, that is,
\begin{align*}
X_t:=r_t-\hat{\mathbb{E}}[r_t]=\int_0^te^{-\alpha(t-u)}dB_u.
\end{align*}
By Proposition \ref{solution to short rate dynamics}, we know that $X$ solves the $G$-stochastic differential equation
\begin{align*}
X_t=-\int_0^t\alpha X_udu+B_t.
\end{align*}
Then the process $Y=(Y_t)_{0\leq t\leq T}$, defined by $Y_t:=\hat{\mathbb{E}}_t[X_T^2]$, is given by $Y_t=u(t,X_t)$ \citep*[Theorems 4.4, 4.5]{hujipengsong2014}, where the function $u:[0,T]\times\mathbb{R}\rightarrow\mathbb{R}$ is the unique viscosity solution to the nonlinear partial differential equation
\begin{align*}
\partial_tu+G(\partial_x^2u)-\alpha x\partial_xu=0,\quad u(T,x)=x^2.
\end{align*}
One can verify that the solution to the nonlinear partial differential equation is given by
\begin{align*}
u(t,x)=\tfrac{\overline{\sigma}^2}{2\alpha}(1-e^{-2\alpha(T-t)})+e^{-2\alpha(T-t)}x^2.
\end{align*}
This proves the first equality in \eqref{variance of the short rate}. The second follows by the same procedure.
\end{proof}

\section{Related Bond Market}\label{related bond market}
The corresponding bond market consists of a money-market account and zero-coupon bonds for all possible maturities. First of all, we fix a finite time $\tau<\infty$ and suppose that all trading takes place within the finite time horizon $[0,\tau]$. The market consists of the following investment opportunities. The first one is to invest in the money-market account, which grows by the short rate $r$. The money-market account is a process denoted by $M=(M_t)_{0\leq t\leq\tau}$ and is given by
\begin{align*}
M_t:=\exp\Big(\int_0^tr_sds\Big).
\end{align*}
In addition to the money-market account, the market offers zero-coupon bonds for all maturities within the time horizon. For $T\leq\tau$, the price of a bond with maturity $T$ at time $t$ is denoted by $P_t(T)$ for $t\leq T$. The bond has a terminal payoff of $1$, that is,
\begin{align*}
P_T(T)=1
\end{align*}
for all $T$. Henceforth, we use the money-market account as a num\'eraire. That means, we restrict to the discounted bonds $\tilde{P}(T)=(\tilde{P}_t(T))_{0\leq t\leq T}$ for $T\leq\tau$, defined by
\begin{align*}
\tilde{P}_t(T):=M_t^{-1}P_t(T).
\end{align*}
We assume that, for all $T$, the discounted bond $\tilde{P}(T)$ is a diffusion process, i.e.,
\begin{align*}
\tilde{P}_t(T)=\tilde{P}_0(T)+\int_0^t\alpha_u(T)du+\int_0^t\beta_u(T)dB_u+\int_0^t\gamma_u(T)d\langle B\rangle_u
\end{align*}
for processes $\alpha(T)=(\alpha_t(T))_{0\leq t\leq T}$, $\beta(T)=(\beta_t(T))_{0\leq t\leq T}$, and $\gamma(T)=(\gamma_t(T))_{0\leq t\leq T}$ in $M_G^2(0,T)$. This is a technical assumption to make the following definition work. The assumption is satisfied in all of the succeeding scenarios.
\par The agents can participate in the market by choosing a trading strategy to create a portfolio. Choosing a market strategy means that they can select a finite number of discounted bonds they want to trade and decide on how much of them they want to buy or sell at each time within the time horizon. The value of the related portfolio is the integral of the market strategy with respect to the price processes. That means, we implicitly assume that the trading strategy is self-financing.
\begin{df}\label{market strategy}
An admissible market strategy $(\pi,T)$ is a couple consisting of a bounded process $\pi=(\pi_t^1,...,\pi_t^n)_{0\leq t\leq\tau}$ in $M_G^2(0,\tau;\mathbb{R}^n)$ and a vector $T=(T_1,...,T_n)\in[0,\tau]^n$ for some $n\in\mathbb{N}$. The corresponding portfolio value at terminal time is defined by
\begin{align*}
\tilde{v}_\tau(\pi,T):=\sum_{i=1}^n\int_0^{T_i}\pi_t^id\tilde{P}_t(T_i).
\end{align*}
\end{df}
\noindent The restriction to trading a finite number of discounted bonds could be generalized by using methods from large financial markets \citep*{kleinschmidtteichmann2016} or by allowing for continuous trading within the set of all maturities \citep*{bjorkdimasikabanovrunggaldier1997}. Here we restrict to trading finitely many discounted bonds, since such a generalization is not the objective of the model.
\par We use a quasi-sure notion of arbitrage. The classical definition of arbitrage depends on the underlying probability measure of the model. Since we are dealing with more than one measure in the presence of volatility uncertainty, we need to consider a definition slightly different from the classical one. The following definition of arbitrage corresponds to the one that is commonly used in the literature on robust finance \citep*[for example]{bouchardnutz2015}.
\begin{df}\label{arbitrage}
An admissible market strategy $(\pi,T)$ is called arbitrage strategy if
\begin{align*}
\tilde{v}_\tau(\pi,T)\geq0\quad\text{quasi-surely},\quad P\big(\tilde{v}_\tau(\pi,T)>0\big)>0\quad\text{for at least one }P\in\mathcal{P}.
\end{align*}
Moreover, we say that the bond market is arbitrage-free if there is no arbitrage strategy.
\end{df}
\noindent This is a weaker version than requiring that the strategy has to be an arbitrage in the classical sense with respect to all measures. The difference is that the probability of a strictly positive win does not have to be strictly positive under each measure.

\section{Martingale Modeling}\label{martingale modeling}
Most short rate models use a martingale modeling approach to ensure that the related bond market is arbitrage-free. A standard result in mathematical finance is that the market is arbitrage-free if and only if the traded quantities on the market are martingales under a measure equivalent to the real world measure, termed fundamental theorem of asset pricing. The common practice in short rate models is martingale modeling, since bond markets are incomplete. Incomplete means that there is not a unique martingale measure but many of them. Thus, one usually supposes that the short rate satisfies certain dynamics under a given martingale measure. Then the bond prices are chosen such that the discounted bonds are martingales under the exogenously given martingale measure in order to exclude arbitrage.
\par In the presence of volatility uncertainty, martingale modeling requires that the discounted bonds are symmetric $G$-martingales under $\hat{\mathbb{E}}$. If there is volatility uncertainty, the set of beliefs contains mutually singular measures. Hence, there is no dominating measure for the set of beliefs, which implies that it is not possible to find a single martingale measure equivalent to all measures in the set of beliefs. A fundamental theorem of asset pricing under a possibly nondominated set of beliefs was established by \citet*{bouchardnutz2015}, for the discrete-time case, and \citet*{biaginibouchardkardarasnutz2017}, for the continuous-time case. Roughly speaking, the theorem says that the absence of arbitrage is equivalent to the existence of a set of martingale measures which is in some sense equivalent to the set of beliefs. That means, the price processes have to be martingales under each measure in the equivalent set of measures. Therefore, if we want to follow a martingale modeling approach in the presence of volatility uncertainty, we need to assume that our set of beliefs is a set of exogenously given martingale measures. Then we need to choose the bond prices such that the discounted bonds are martingales under each measure in the set of beliefs. Being a martingale under each measure in the set of beliefs is equivalent to being a symmetric $G$-martingale under $\hat{\mathbb{E}}$. The sufficiency of this martingale modeling approach for the absence of arbitrage is shown by the following proposition.
\begin{prp}\label{no-arbitrage if the discounted bonds are symmetric G-martingales}
The bond market is arbitrage-free if the discounted bond $\tilde{P}(T)$ is a symmetric $G$-martingale under $\hat{\mathbb{E}}$ for all $T$.
\end{prp}
\begin{proof}
We suppose that there exists an arbitrage strategy $(\pi,T)$ and show that this leads to a contradiction. By Definition \ref{arbitrage}, it holds $\tilde{v}_\tau(\pi,T)\geq0$. Hence, we know that $\vert\tilde{v}_\tau(\pi,T)\vert=\tilde{v}_\tau(\pi,T)$, which implies
\begin{align*}
\hat{\mathbb{E}}[\vert\tilde{v}_\tau(\pi,T)\vert]=\hat{\mathbb{E}}[\tilde{v}_\tau(\pi,T)].
\end{align*}
Using Definition \ref{market strategy} and the sublinearity of $\hat{\mathbb{E}}$, we obtain
\begin{align*}
\hat{\mathbb{E}}[\tilde{v}_\tau(\pi,T)]\leq\sum_{i=1}^n\hat{\mathbb{E}}\Big[\int_0^{T_i}\pi_t^id\tilde{P}_t(T_i)\Big].
\end{align*}
By the representation theorem for symmetric $G$-martingales \citep*[Theorem 4.8]{song2011}, for all $T$, there exists a process $H(T)=(H_t(T))_{0\leq t\leq T}$ in $M_G^2(0,T)$ such that
\begin{align*}
\tilde{P}_t(T)=\tilde{P}_0(T)+\int_0^tH_u(T)dB_u.
\end{align*}
Since $\pi^i$ is a bounded process in $M_G^2(0,\tau)$, we have $\pi^iH(T_i)\in M_G^2(0,T_i)$ for all $i$. Thus,
\begin{align*}
\hat{\mathbb{E}}\Big[\int_0^{T_i}\pi_t^id\tilde{P}_t(T_i)\Big]=\hat{\mathbb{E}}\Big[\int_0^{T_i}\pi_t^iH_t(T_i)dB_t\Big]=0
\end{align*}
for all $i$. Combining the previous steps, we get $\tilde{v}_\tau(\pi,T)=0$, which is a contradiction to
\begin{align*}
P\big(\tilde{v}_\tau(\pi,T)>0\big)>0\quad\text{for at least one }P\in\mathcal{P}.
\end{align*}
Therefore, there is no arbitrage strategy.
\end{proof}
\par Unfortunately, we can show that the martingale modeling approach in the Hull-White model does not work in the presence of volatility uncertainty. Martingale modeling only works in the classical case when there is no uncertainty about the volatility. In that case the bond prices are obviously given by the bond prices from the classical Hull-White model.
\begin{thm}\label{martingale modeling iff no volatility uncertainty}
The discounted bond $\tilde{P}(T)$ is a symmetric $G$-martingale under $\hat{\mathbb{E}}$ if and only if $\overline{\sigma}=\underline{\sigma}$ and the bond price is given by
\begin{align}\label{bond price from the Hull-White model}
P_t(T)=\exp\big(A^{\overline{\sigma}}(t,T)-B(t,T)r_t\big)
\end{align}
for all $t$, where $A^\sigma,B:[0,\tau]\times[0,\tau]\rightarrow\mathbb{R}$, for $\sigma>0$, are defined by
\begin{subequations}
\begin{align}
A^\sigma(t,T):={}&\int_t^T\big(\tfrac{1}{2}\sigma^2B(s,T)^2-\mu(s)B(s,T)\big)ds,\label{definition of A sigma}
\\B(t,T):={}&\tfrac{1}{\alpha}(1-e^{-\alpha(T-t)}),\label{definition of B}
\end{align}
\end{subequations}
respectively.
\end{thm}
\begin{proof}
First, let us suppose that the discounted bond $\tilde{P}(T)$ is a symmetric $G$-martingale under $\hat{\mathbb{E}}$. We show that the expectation of the discount factor is the same under each measure in the set of beliefs. The definition of symmetric $G$-martingales and the terminal condition of the bond implies
\begin{align*}
\tilde{P}_t(T)=\hat{\mathbb{E}}_t[\tilde{P}(T,T)]&{}=\hat{\mathbb{E}}_t[M_T^{-1}],
\\\tilde{P}_t(T)=-\hat{\mathbb{E}}_t[-\tilde{P}(T,T)]&{}=-\hat{\mathbb{E}}_t[-M_T^{-1}]
\end{align*}
for all $t$. Combining the previous equations and setting $t=0$ yields
\begin{align}\label{supinfeq}
\sup_{P\in\mathcal{P}}\mathbb{E}_P[M_T^{-1}]=\inf_{P\in\mathcal{P}}\mathbb{E}_P[M_T^{-1}],
\end{align}
which in turn implies that the expectation of $M_T^{-1}$ is the same under each measure.
\par Now we use the expression for the bond prices from the classical Hull-White model to show that \eqref{supinfeq} implies $\overline{\sigma}=\underline{\sigma}$ and \eqref{bond price from the Hull-White model}. Let us consider the measures $P^{\overline{\sigma}},P^{\underline{\sigma}}\in\mathcal{P}$ induced by the highest and the lowest possible volatility, respectively. The expectations of $M_T^{-1}$ under $P^{\overline{\sigma}}$ and $P^{\underline{\sigma}}$ are given by
\begin{align*}
\mathbb{E}_{P^{\overline{\sigma}}}[M_T^{-1}]={}&\exp\big(A^{\overline{\sigma}}(0,T)-B(0,T)r_0\big),
\\\mathbb{E}_{P^{\underline{\sigma}}}[M_T^{-1}]={}&\exp\big(A^{\underline{\sigma}}(0,T)-B(0,T)r_0\big),
\end{align*}
respectively \citep*[Subsection 22.4.4]{bjork2004}. By \eqref{supinfeq}, the latter expressions are equal. From \eqref{definition of A sigma} and \eqref{definition of B}, we see that this only holds if $\overline{\sigma}=\underline{\sigma}$. Since $\overline{\sigma}=\underline{\sigma}$ implies $\mathcal{P}=\{P^{\overline{\sigma}}\}$, we are back in the classical case without volatility uncertainty. In that case, the bond price is given by \eqref{bond price from the Hull-White model}.
\par Next, let us suppose that $\overline{\sigma}=\underline{\sigma}$ and the bond price is determined by \eqref{bond price from the Hull-White model}. Then we are again back in the classical case without volatility uncertainty and the discounted bond is clearly a martingale under $P^{\overline{\sigma}}$. Since $\mathcal{P}=\{P^{\overline{\sigma}}\}$, the discounted bond is also a symmetric $G$-martingale.
\end{proof}

\section{Equivalent Sublinear Expectations}\label{equivalent sublinear expectations}
In order to find an arbitrage-free term structure, we consider the following type of sublinear expectations defined by a $G$-backward stochastic differential equation. Let $\lambda=(\lambda_t)_{0\leq t\leq\tau}$ be a bounded process in $M_G^p(0,\tau)$ for some $p>1$. For $\xi\in L_G^p(\Omega_\tau)$ with $p>1$, we define the sublinear expectation $\bar{\mathbb{E}}$ by $\bar{\mathbb{E}}_t[\xi]:=Y_t^\xi$, where $Y^\xi=(Y_t^\xi)_{0\leq t\leq\tau}$ solves the $G$-backward stochastic differential equation
\begin{align*}
Y_t^\xi=\xi+\int_t^\tau\lambda_uZ_udu-\int_t^\tau Z_udB_u-(K_\tau-K_t).
\end{align*}
Then $\bar{\mathbb{E}}$ is a time consistent sublinear expectation \citep*[Theorem 5.1]{hujipengsong2014}. The reader may refer to the paper of \citet*{hujipengsong2014} for all details related to $G$-backward stochastic differential equations.
\par We can show that a sublinear expectation of the above kind is equivalent to the initial sublinear expectation in the sense that the null spaces induced by the natural norms related to both sublinear expectations are the same.
\begin{lem}\label{equivalence between sublinear expectations}
For $\xi\in L_G^p(\Omega_\tau)$ with $p>1$, $\xi=0$ if and only if $\bar{\mathbb{E}}[\vert\xi\vert]=0$.
\end{lem}
\begin{proof}
Before we show the assertion, we explicitly solve the $G$-backward stochastic differential equation defining $\bar{\mathbb{E}}$. For this purpose, we consider the extended $\tilde{G}$-expectation space $(\tilde{\Omega}_\tau,L_{\tilde{G}}^1(\tilde{\Omega}_\tau),\tilde{\mathbb{E}})$ with the canonical process $(B,\tilde{B})=(B_t,\tilde{B}_t)_{t\geq0}$, where $\tilde{\Omega}_\tau=C_0^2([0,\tau])$ and the generator $\tilde{G}:\mathbb{S}^2\rightarrow\mathbb{R}$ is given by
\begin{align*}
\tilde{G}(A)=\tfrac{1}{2}\sup_{\sigma\in[\underline{\sigma}^2,\overline{\sigma}^2]}\text{tr}\Big(\begin{pmatrix}\sigma&1\\1&\sigma^{-1}\end{pmatrix}A\Big).
\end{align*}
$C_0^2([0,\tau])$ and $\mathbb{S}^2$ denote the space of all $\mathbb{R}^2$-valued continuous paths on $[0,\tau]$ starting in $0$ and the space of all symmetric $2\times2$ matrices, respectively. Then $Y^\xi$ is given by
\begin{align*}
Y_t^\xi=\mathcal{E}_t^{-1}\tilde{\mathbb{E}}_t[\mathcal{E}_\tau\xi]
\end{align*}
\citep*[Theorem 3.2]{hujipengsong2014}, where the process $\mathcal{E}=(\mathcal{E}_t)_{0\leq t\leq\tau}$ is defined by
\begin{align*}
\mathcal{E}_t:=\exp\Big(\int_0^t\lambda_ud\tilde{B}_u-\tfrac{1}{2}\int_0^t\lambda_u^2d\langle\tilde{B}\rangle_u\Big).
\end{align*}
\par Now we show the assertion by representing $\tilde{\mathbb{E}}$ and $\bar{\mathbb{E}}$ as an upper expectation of a family of probability measures. It suffices to show that $\tilde{\mathbb{E}}[\vert\xi\vert]=0$ if and only if $\bar{\mathbb{E}}[\vert\xi\vert]=0$ for $\xi\in L_G^p(\Omega_\tau)$ with $p>1$, since we know that $\tilde{\mathbb{E}}[\xi]=\hat{\mathbb{E}}[\xi]$ for all $\xi\in L_G^1(\Omega)$. As in Section \ref{short rate dynamics}, we can construct a family $\tilde{\mathcal{P}}$ of probability measures on $(\tilde{\Omega}_\tau,\mathcal{B}(\tilde{\Omega}_\tau))$ such that
\begin{align*}
\tilde{\mathbb{E}}[\xi]=\sup_{\tilde{P}\in\tilde{\mathcal{P}}}\mathbb{E}_{\tilde{P}}[\xi]
\end{align*}
for all $\xi\in L_{\tilde{G}}^1(\tilde{\Omega}_\tau)$. Moreover, $\mathcal{E}$ solves the $G$-stochastic differential equation
\begin{align*}
\mathcal{E}_t=1+\int_0^t\lambda_u\mathcal{E}_ud\tilde{B}_u.
\end{align*}
This implies that $\mathcal{E}$ is a symmetric $G$-martingale, satisfying $\tilde{\mathbb{E}}[\mathcal{E}_\tau]=1$. Thus, for $\tilde{P}\in\tilde{\mathcal{P}}$, we can define a probability measure on $(\tilde{\Omega}_\tau,\mathcal{B}(\tilde{\Omega}_\tau))$ by $Q(\tilde{P}):=\mathcal{E}_\tau\cdot\tilde{P}$. Since $\mathcal{E}_\tau>0$ $\tilde{\mathcal{P}}$-quasi-surely, we know that $Q(\tilde{P})\sim\tilde{P}$. If we now define $\mathcal{Q}:=\{Q(\tilde{P})\vert\tilde{P}\in\tilde{\mathcal{P}}\}$, we obtain
\begin{align*}
\bar{\mathbb{E}}[\xi]=\sup_{Q\in\mathcal{Q}}\mathbb{E}_Q[\xi]
\end{align*}
for all $\xi\in L_G^p(\Omega_\tau)$. Since $\mathcal{Q}$ consists of equivalent measures, we get that $\xi=0$ $\tilde{\mathcal{P}}$-quasi-surely if and only if $\xi=0$ $\mathcal{Q}$-quasi surely. Hence, the proof is complete.
\end{proof}
\par As a consequence, we can show that there is no abitrage on the bond market if there exists an equivalent sublinear expectation of the above kind under which the discounted bonds are symmetric $G$-martingales.
\begin{prp}\label{no-arbitrage if the discounted bonds are symmetric G-martingales under the equivalent sublinear expectation}
The bond market is arbitrage-free if the discounted bond $\tilde{P}(T)$ is a symmetric $G$-martingale under $\bar{\mathbb{E}}$ for all $T$.
\end{prp}
\begin{proof}
We proceed as in the proof of Proposition \ref{no-arbitrage if the discounted bonds are symmetric G-martingales} by using Lemma \ref{equivalence between sublinear expectations} and the Girsanov transformation for $G$-Brownian motion from \citet*{hujipengsong2014}. Let us suppose that there exists an arbitrage strategy $(\pi,T)$. By Definition \ref{arbitrage}, it holds $\tilde{v}_\tau(\pi,T)\geq0$, which implies $\vert\tilde{v}_\tau(\pi,T)\vert=\tilde{v}_\tau(\pi,T)$. By Lemma \ref{equivalence between sublinear expectations}, we have
\begin{align*}
\bar{\mathbb{E}}[\vert\tilde{v}_\tau(\pi,T)\vert]=\bar{\mathbb{E}}[\tilde{v}_\tau(\pi,T)].
\end{align*}
Using Definition \ref{market strategy} and the sublinearity of $\bar{\mathbb{E}}$, we obtain
\begin{align*}
\bar{\mathbb{E}}[\tilde{v}_\tau(\pi,T)]\leq\sum_{i=1}^n\bar{\mathbb{E}}\Big[\int_0^{T_i}\pi_t^id\tilde{P}_t(T_i)\Big].
\end{align*}
By the Girsanov transformation for $G$-Brownian motion \citep*[Theorem 5.2]{hujipengsong2014}, the process $\bar{B}=(\bar{B}_t)_{0\leq t\leq\tau}$, defined by
\begin{align*}
\bar{B}_t:=B_t-\int_0^t\lambda_udu,
\end{align*}
is a $G$-Brownian motion under $\bar{\mathbb{E}}$. Since $\tilde{P}(T)$ is a symmetric $G$-martingale under $\bar{\mathbb{E}}$, for all $T$, there exists a process $H(T)=(H_t(T))_{0\leq t\leq T}$ in $M_G^2(0,T)$ such that
\begin{align*}
\tilde{P}_t(T)=\tilde{P}_0(T)+\int_0^tH_u(T)d\bar{B}_u.
\end{align*}
Thus, as in the proof of Proposition \ref{no-arbitrage if the discounted bonds are symmetric G-martingales}, we have
\begin{align*}
\bar{\mathbb{E}}\Big[\int_0^{T_i}\pi_t^id\tilde{P}_t(T_i)\Big]=\bar{\mathbb{E}}\Big[\int_0^{T_i}\pi_t^iH_t(T_i)d\bar{B}_t\Big]=0
\end{align*}
for all $i$. Combining the previous steps, we get $\tilde{v}_\tau(\pi,T)=0$ by Lemma \ref{equivalence between sublinear expectations}, which is a contradiction to
\begin{align*}
P\big(\tilde{v}_\tau(\pi,T)>0\big)>0\quad\text{for at least one }P\in\mathcal{P}.
\end{align*}
Therefore, there is no arbitrage strategy.
\end{proof}

\section{Arbitrage-Free Term Structure}\label{arbitrage-free term structure}
There exists an equivalent sublinear expectation of the above kind under which the discounted bonds are symmetric $G$-martingales. We define the process $q=(q_t)_{0\leq t\leq\tau}$ by
\begin{align*}
q_t:=\int_0^te^{-2\alpha(t-u)}d\langle B\rangle_u.
\end{align*}
By applying It\^o's formula for $G$-Brownian motion, we observe that $q$ satisfies
\begin{align*}
q_t=\langle B\rangle_t-\int_0^t2\alpha q_udu.
\end{align*}
If we use the process $q$ to define an equivalent sublinear expectation as in Section \ref{equivalent sublinear expectations}, we obtain a sublinear expectation under which there is a unique expression for the bond prices such that the discounted bonds are symmetric $G$-martingales. A justification for choosing the process $q$ is given below.
\begin{thm}\label{symmetric G-martingale under the equivalent sublinear expectation}
Let $\lambda=q$. Then the discounted bond $\tilde{P}(T)$ is a symmetric $G$-martingale under $\bar{\mathbb{E}}$ if and only if the bond price is given by
\begin{align}\label{arbitrage-free bond prices}
P_t(T)=\exp\big(A(t,T)-B(t,T)r_t-\tfrac{1}{2}B(t,T)^2q_t\big)
\end{align}
for all $t$, where $A,B:[0,\tau]\times[0,\tau]\rightarrow\mathbb{R}$ are defined by
\begin{align}\label{definition of A}
A(t,T):=-\int_t^TB(s,T)\mu(s)ds
\end{align}
and \eqref{definition of B}, respectively.
\end{thm}
\begin{proof}
First of all, we show that the process $X=(X_t)_{0\leq t\leq T}$, defined by
\begin{align*}
X_t:=\exp\Big(A(t,T)-B(t,T)r_t-\tfrac{1}{2}B(t,T)^2q_t-\int_0^tr_sds\Big),
\end{align*}
is a symmetric $G$-martingale under $\hat{\mathbb{E}}$. Applying It\^o's formula for $G$-Brownian motion to $X$ leads to the dynamics
\begin{align*}
X_t=X_0+\int_0^t\Delta_uX_udu-\int_0^tB(u,T)X_udB_u+\int_0^t\Delta_u'X_ud\langle B\rangle_u,
\end{align*}
where the drift terms $\Delta=(\Delta_t)_{0\leq t\leq T}$ and $\Delta'=(\Delta'_t)_{0\leq t\leq T}$ are given by
\begin{align*}
\Delta_t:={}&\partial_tA(t,T)-\partial_tB(t,T)r_t-B(t,T)\partial_tB(t,T)q_t
\\&{}-B(t,T)(\mu(t)-\alpha r_t)+B(t,T)^2\alpha q_t-r_t
\\={}&\big(\partial_tA(t,T)-\mu(t)B(t,T)\big)-\big(\partial_tB(t,T)-\alpha B(t,T)+1\big)r_t
\\&-B(t,T)\big(\partial_tB(t,T)-\alpha B(t,T)\big)q_t,
\\\Delta_t':={}&-\tfrac{1}{2}B(t,T)^2+\tfrac{1}{2}B(t,T)^2=0,
\end{align*}
respectively. The functions $A$ and $B$ satisfy
\begin{align*}
\partial_tA(t,T)={}&\mu(t)B(t,T),
\\\partial_tB(t,T)={}&\alpha B(t,T)-1,
\end{align*}
respectively. Thus, we get
\begin{align*}
X_t=X_0+\int_0^tB(u,T)X_uq_udu-\int_0^tB(u,T)X_udB_u.
\end{align*}
Since the previous equation is a linear $G$-stochastic differential equation with bounded coefficients, it has a unique solution, which is in $M_G^2(0,T)$. Therefore, $X\in M_G^2(0,T)$, which implies $X_t\in L_G^2(\Omega_t)$ for all $t$. By the Girsanov transformation for $G$-Brownian motion, the process $\bar{B}=(\bar{B}_t)_{0\leq t\leq\tau}$, defined by
\begin{align*}
\bar{B}_t:=B_t-\int_0^tq_udu,
\end{align*}
is a $G$-Brownian motion under $\bar{\mathbb{E}}$. Hence, $X$ is a symmetric $G$-martingale under $\bar{\mathbb{E}}$.
\par Using the first step, we now prove the assertion. If $\tilde{P}(T)$ is a symmetric $G$-martingale under $\bar{\mathbb{E}}$, for all $t$, it holds
\begin{align*}
\tilde{P}_t(T)=\bar{\mathbb{E}}_t[\tilde{P}_T(T)]=\bar{\mathbb{E}}_t[M_T^{-1}].
\end{align*}
Since $X$ is also a symmetric $G$-martingale under $\bar{\mathbb{E}}$ and $A(T,T)=0=B(T,T)$, we get
\begin{align*}
X_t=\bar{\mathbb{E}}_t[X_T]=\bar{\mathbb{E}}_t[M_T^{-1}]
\end{align*}
for all $t$. Thus, we have $\tilde{P}_t(T)=X_t$ for all $t$, which is equivalent to \eqref{arbitrage-free bond prices}. Conversly, if \eqref{arbitrage-free bond prices} holds, we get $\tilde{P}_t(T)=X_t$ for all $t$ and hence, $\tilde{P}(T)$ is a symmetric $G$-martingale by the first step of the proof.
\end{proof}
\par We use the process $q$ to obtain an arbitrage-free term structure, since it serves as an adjustment factor for the uncertain volatility. The proof of Theorem \ref{martingale modeling iff no volatility uncertainty} shows that the discounted bond cannot be a symmetric $G$-martingale under $\hat{\mathbb{E}}$ in the presence of volatility uncertainty, since the expectation of the discount factor is not the same for every measure in the set of beliefs. The expectation differs among the measures in $\mathcal{P}$, since the short rate has a different variance under each of them, as Theorem \ref{mean and variance of the short rate} shows. So in order to unify the expectation of the discount factor under each measure, we need to adjust the short rate by the uncertainty about its variance. The following identity shows that the process $q$ is a suitable adjustment factor, since it contains the same information as the variance of the short rate. By Proposition \ref{solution to short rate dynamics} and a standard property of integrals with respect to $G$-Brownian motion \citep*[Proposition 3.4.5]{peng2019}, we have
\begin{align*}
\hat{\mathbb{E}}[(r_t-\hat{\mathbb{E}}[r_t])^2]=\hat{\mathbb{E}}\Big[\Big(\int_0^te^{-\alpha(t-u)}dB_u\Big)^2\Big]=\hat{\mathbb{E}}\Big[\int_0^te^{-2\alpha(t-u)}d\langle B\rangle_u\Big]=\hat{\mathbb{E}}[q_t].
\end{align*}
Therefore, we set $\lambda=q$ and use the Girsanov transformation for $G$-Brownian motion to adjust the short rate by its variance. Then the short rate evolves according to the dynamics
\begin{align*}
r_t=r_0+\int_0^t\big(\mu(u)-\alpha r_u+q_u\big)du+\bar{B}_t.
\end{align*}
Another important observation, which can be deduced from the dynamics of $q$, is that the process $q$ mean reverts twice as fast as the short rate towards the quadratic variation of the $G$-Brownian motion and thus, towards the quadratic variation of the short rate. So the process always adjusts towards the correct belief about the volatility, which is unknown beforehand.
\par From an economic point of view, setting $\lambda=q$ is reasonable as well. In the proof of Theorem \ref{symmetric G-martingale under the equivalent sublinear expectation} we see that the instantaneous excess return of a zero-coupon bond with maturity $T$ over the money-market account at time $t$ is given by $B(t,T)q_t$. Dividing by the diffusion coefficient $-B(t,T)$, we obtain the market price of risk $-q_t$. In general, the market price of risk measures how much better we are doing with a bond compared to investing in the money-market account per one unit of risk. Since $q$ is positive, we use a negative market price of risk. This is appropriate, because the bonds are not risky in this model. They have a certain payoff of $1$ at the maturity, i.e., there is no default risk. On the other hand, investing in the money-market account is risky, since the short rate is stochastic and uncertain. Hence, we use a process representing the variance of the short rate to measure the risk and the uncertainty of the money-market account. So one may also refer to $-q$ as the market price of uncertainty.
\par In order to compare the bond prices from Theorem \ref{symmetric G-martingale under the equivalent sublinear expectation} with the prices from the traditional model, we derive an adjustment factor, linking both expressions. We denote the bond price from the traditional Hull-White model with constant volatility $\sigma$ by $P_t^\sigma(T)$, which is defined by
\begin{align*}
P_t^\sigma(T):=\exp\big(A^\sigma(t,T)-B(t,T)r_t\big),
\end{align*}
where $A^\sigma(t,T)$ and $B(t,T)$ are defined by \eqref{definition of A sigma} and \eqref{definition of B}, respectively. The bond price of the present model $P_t(T)$ is given by \eqref{arbitrage-free bond prices}. Then we have
\begin{align*}
\frac{P_t(T)}{P_t^\sigma(T)}=\exp\Big(-\int_t^T\tfrac{1}{2}\sigma^2B(s,T)^2ds-\tfrac{1}{2}B(t,T)^2q_t\Big).
\end{align*}
The expression on the right-hand side represents an adjustment factor, which we can use to migrate from the traditional model to the model with volatility uncertainty.
\par Examining the adjustment factor, we note the following differences between the traditional and the present model. Since the adjustment factor is less than one, the bond prices in the present model are less than the prices in the classical model without volatility uncertainty. Moreover, we see that the squared term, depending on the volatility $\sigma$, is missing in $P_t(T)$. Instead, we have an additional term in $P_t(T)$, depending on the market price of uncertainty. Thus, the prices are independent of the volatility as well as the bounds for the volatility, which is the case in most of the models dealing with pricing under volatility uncertainty. It also implies that the bond price at the initial time corresponds to the price in the deterministic version of the Hull-White model without white noise, since the additional part in the exponential vanishes at the initial time. Though, this also applies to the standard Hull-White model after fitting it to the initial yield curve. In contrast to classical affine models, the bond price is now affine with respect to the short rate and the market price of uncertainty. The affine structure is similar to short rate models with a stochastic volatility \citep*{fongvasicek1991,longstaffschwartz1992}. However, the additional factor in the bond price is not the volatility but a process adjusting towards the current value of the quadratic variation of the short rate. A surprisingly similar structure can be found in the short rate model from \citet*{casassuscollindufresnegoldstein2005}, displaying unspanned stochastic volatility.
\par The most important implications of the prices in this model are as follows. Primarily, we manage to obtain a term structure which is robust with respect to the volatility itself and the bounds for the volatility. So we do neither have to estimate the volatility of the short rate in the future nor its bounds. Admittedly, there is a price we have to pay for this. We have to specify the market price of uncertainty, which appears in the bond prices. The market price of uncertainty depends on the past evolution of the quadratic variation of the $G$-Brownian motion, which corresponds to the quadratic variation of the short rate in the Hull-White model. The past evolution of the quadratic variation of the short rate is observable and can be inferred from market data. Alternatively, one could also estimate the variance of the short rate as an approximation for the market price of uncertainty. Moreover, in accordance to the short rate model of \citet*{casassuscollindufresnegoldstein2005}, the bond prices are completely unaffected by the structure of the volatility. The bounds of the volatility probably enter the model when it is used for pricing derivatives on bonds, that is, nonlinear contracts. Therefore, the model, despite its simple structure, could be a contribution to the literature on unspanned stochastic volatility, which was introduced by \citet*{collindufresnegoldstein2002}. A detailed discussion and the pricing of derivatives is, however, left for a subsequent study.

\section{Yield Curve Fitting}\label{yield curve fitting}
As in the classical Hull-White model, we can use the time dependent mean reversion level to fit the theoretical bond prices to an initially observable term structure. We introduce the following notions and assumptions, which are common in term structure models. Let us assume that there is an initial forward curve $f_0^*:[0,\tau]\rightarrow\mathbb{R}$, which is observed on the market. We assume that the initial forward curve $f_0^*$ is differentiable and satisfies $f_0^*(0)=r_0$. For $T\leq\tau$, the theoretical forward rate of the model is denoted by $f_t(T)$ for $t\leq T$ and defined by
\begin{align*}
f_t(T):=-\partial_T\log P_t(T).
\end{align*}
The following theorem gives a necessary and sufficient condition for the theoretical forward curve matching the observable one at inception, which characterizes the mean reversion level of the short rate.
\begin{thm}\label{characterization of the mean reversion level in terms of an initial yield curve}
Let the bond price be given by \eqref{arbitrage-free bond prices}. Then it holds $f_0^*(T)=f_0(T)$ for all $T$ if and only if the mean reversion level of the short rate, for all $t$, satisfies
\begin{align}\label{mean reversion level after yield curve fitting}
\mu(t)=\alpha f_0^*(t)+\partial_tf_0^*(t).
\end{align}
\end{thm}
\begin{proof}
First, we derive the initial forward rate $f_0(T)$ for $T\leq\tau$ when the bond price is given by \eqref{arbitrage-free bond prices}. Taking the derivative of the logarithm of the bond price at time $0$ and changing the sign, for $T\leq\tau$, we obtain
\begin{align*}
f_0(T)=\int_0^T\mu(t)e^{-\alpha(T-t)}dt+e^{-\alpha T}r_0.
\end{align*}
\par Let us suppose that $f_0^*(T)=f_0(T)$ for all $T$. By the equation from above, we have
\begin{align*}
e^{\alpha T}f_0^*(T)=\int_0^T\mu(t)e^{\alpha t}dt+r_0
\end{align*}
for all $T$. Differentiating the latter equation with respect to $T$ yields
\begin{align*}
\alpha e^{\alpha T}f_0^*(T)+e^{\alpha T}\partial_Tf_0^*(T)=\mu(T)e^{\alpha T}.
\end{align*}
for all $T$. Hence, the mean reversion level satisfies \eqref{mean reversion level after yield curve fitting}.
\par If we suppose that the mean reversion level satisfies \eqref{mean reversion level after yield curve fitting}, we can plug it into the first equation of the proof and check, by reversing the above calculations, that it holds $f_0^*(T)=f_0(T)$ for all $T$.
\end{proof}
\par After fitting the yield curve, the model is consistent with the classical Hull-White model. In general, the model is not consistent with the traditional Hull-White model, since the short rate dynamics and the bond prices differ from the ones in the traditional model, even if there is no volatility uncertainty. The adjusted short rate dynamics are given by
\begin{align*}
r_t=r_0+\int_0^t\big(\mu(u)+q_u-\alpha r_u\big)du+\bar{B}_t,
\end{align*}
which clearly differ from the Hull-White short rate dynamics. The bond prices also differ from the ones obtained in the Hull-White model, as the comparison from the previous section shows. However, the model becomes consistent with the classical one after fitting the model to the yield curve. After inserting \eqref{mean reversion level after yield curve fitting} and the definition of $q$ in the short rate dynamics, one can check that these are the same dynamics as in the fitted Hull-White model if there is no volatility uncertainty, i.e., if $\overline{\sigma}=\sigma=\underline{\sigma}$ \citep*[Subsection 3.3.1]{brigomercurio2001}. Furthermore, inserting \eqref{mean reversion level after yield curve fitting} in $A(t,T)$, which is defined in \eqref{definition of A}, and performing some calculations, yields
\begin{align*}
A(t,T)=-\int_t^Tf_0^*(s)ds+f_0^*(t)B(t,T).
\end{align*}
Plugging the above expression into \eqref{arbitrage-free bond prices} and dividing by the fitted bond price from the traditional Hull-White model with volatility $\sigma$ \citep*[Subsection 3.3.2]{brigomercurio2001} leads to
\begin{align*}
\frac{P_t(T)}{P_t^\sigma(T)}=\exp\Big(\tfrac{1}{2}B(t,T)^2\big(\tfrac{\sigma^2}{2\alpha}(1-e^{-2\alpha t})-q_t\big)\Big).
\end{align*}
Thus, the adjustment factor is now determined by the difference between the variance of the short rate with constant volatility and the uncertain variance of the short rate with volatility uncertainty. Hence, the adjustment factor is equal to $1$ if $\overline{\sigma}=\sigma=\underline{\sigma}$.
\par The consistency with the classical Hull-White model further justifies the choice of $q$ as the market price of uncertainty. The discussion from above shows that the adjustment factor $q$ appearing in the short rate dynamics is actually included in the dynamics of the classical model after fitting the theoretical prices to the observed ones. Hence, the process $q$ naturally appears in the risk-neutral dynamics of the short rate, which gives another justification for choosing the market price of uncertainty in this particular way. The interesting thing, however, is that in the classical model, this expression is used for yield curve fitting, whereas in this model, the expression is needed in order to have an arbitrage-free model.
\par In order to completely calibrate the model, one has to establish a robust estimation procedure for the mean reversion speed. Theorem \ref{characterization of the mean reversion level in terms of an initial yield curve} characterizes the mean reversion level in terms of an initially observable term structure. However, the term structure still involves a parameter: the mean reversion speed $\alpha$. A typical approach in short rate models to estimate parameters is to use a maximum lieklihood approach. The maximum likelihood approach heavily relies on the probabilistic law of the short rate. In the presence of volatility uncertainty, there is a family of possible probabilistic laws for the short rate and we are uncertain about which one is correct. Therefore, one has to use a robust approach to calibrate the model instead of the classical maximum likelihood approach.

\section{Multifactor Extension}\label{multifactor extension}
The previous model can be generalized to a model driven by multiple factors. For this purpose, we consider a $d$-dimensional $G$-Brownian motion $B=(B_t^1,...,B_t^d)_{t\geq0}$. That is, $B^i=(B_t^i)_{t\geq0}$ is a $1$-dimensional $G$-Brownian motion for all $i$. A $d$-dimensional $G$-Brownian motion can be constructed as in Section \ref{short rate dynamics} by replacing $C_0(\mathbb{R}_+)$ and $[\underline{\sigma},\overline{\sigma}]$ with $C_0^d(\mathbb{R}_+)$, the space of all $\mathbb{R}^d$-valued continuous paths on $\mathbb{R}_+$ starting in $0$, and a bounded, closed, and convex subset $\Sigma\subset\mathbb{R}^{d\times d}$, respectively. Hence, we allow for a possible uncertain correlation between the risk factors. The short rate process $r$ is defined by
\begin{align*}
r_t:=\mu(t)+\sum_{i=1}^dX_t^i,
\end{align*}
where $\mu:\mathbb{R}_+\rightarrow\mathbb{R}$ is a suitably integrable function and the factor $X^i=(X_t^i)_{t\geq0}$ satisfies
\begin{align*}
X_t^i=-\int_0^t\alpha_iX_u^idu+B_t^i
\end{align*}
for some constant $\alpha_i>0$ for all $i$. The process $X^i$ is given by
\begin{align*}
X_t^i=\int_0^te^{-\alpha_i(t-u)}dB_u^i
\end{align*}
for all $i$ and represents a risk factor that affects the short rate.
\par Such a multifactor extension does not lead to an arbitrage-free term structure. Similar to the discussion in Section \ref{martingale modeling}, we can show that the short rate dynamics from above are not suitable for martingale modeling. As in Theorem \ref{martingale modeling iff no volatility uncertainty}, we can show that the discounted bonds can be symmetric $G$-martingales under $\hat{\mathbb{E}}$ if and only if there is no volatility uncertainty, that is, $\Sigma$ is a singleton. This can be shown, as in the proof of Theorem \ref{martingale modeling iff no volatility uncertainty}, by considering two different beliefs about the volatility, which lead to different bond prices.
\par We consider sublinear expectations defined by a $G$-backward stochastic differential equation and equivalent to the initial sublinear expectation in order to find an arbitrage-free term structure. Let $\lambda=(\lambda_t^1,...,\lambda_t^d)_{0\leq t\leq\tau}$ be a $d$-dimensional bounded process in $M_G^p(0,\tau;\mathbb{R}^d)$ for some $p>1$. For $\xi\in L_G^p(\Omega_\tau)$ with $p>1$, we define the sublinear expectation $\bar{\mathbb{E}}$ by $\bar{\mathbb{E}}_t[\xi]:=Y_t^\xi$, where $Y^\xi=(Y_t^\xi)_{0\leq t\leq\tau}$ solves the $G$-backward stochastic differential equation
\begin{align*}
Y_t^\xi=\xi+\sum_{i=1}^d\int_t^\tau\lambda_u^iZ_u^idu-\sum_{i=1}^d\int_t^\tau Z_u^idB_u^i-(K_\tau-K_t).
\end{align*}
Then we can show, as in Proposition \ref{no-arbitrage if the discounted bonds are symmetric G-martingales under the equivalent sublinear expectation}, that the bond market is arbitrage-free if the discounted bonds are symmetric $G$-martingales under $\bar{\mathbb{E}}$.
\par We obtain an arbitrage-free term structure by considering a particular sublinear expectation of the above form. Let us define the process $q=(q_t^1,...,q_t^d)_{0\leq t\leq\tau}$ by
\begin{align*}
q_t^i:=\sum_{j=1}^dq_t^{ij},
\end{align*}
where $q^{ij}=(q_t^{ij})_{0\leq t\leq\tau}$ is defined by
\begin{align*}
q_t^{ij}:=\int_0^te^{-(\alpha_i+\alpha_j)(t-u)}d\langle B^i,B^j\rangle_u.
\end{align*}
By applying It\^o's formula for $G$-Brownian motion, we know that $q^{ij}$, for all $i,j$, satisfies
\begin{align*}
q_t^{ij}=\langle B^i,B^j\rangle_t-\int_0^t(\alpha_i+\alpha_j)q_u^{ij}du.
\end{align*}
If we use the process $q$ to define a sublinear expectation as above, then there is a unique arbitrage-free term structure.
\begin{thm}\label{symmetric G-martingale under the equivalent sublinear expectation with multiple risk factors}
Let $\lambda=q$. Then the discounted bond $\tilde{P}(T)$ is a symmetric $G$-martingale under $\bar{\mathbb{E}}$ if and only if the bond price is given by
\begin{align*}
P_t(T):=\exp\Big(-\int_t^T\mu(s)ds-\sum_{i=1}^dB_i(t,T)X_t^i-\tfrac{1}{2}\sum_{i,j=1}^dB_i(t,T)B_j(t,T)q_t^{ij}\Big)
\end{align*}
for all $t$, where $B_i:[0,\tau]\times[0,\tau]\rightarrow\mathbb{R}$, for all $i$, is defined by
\begin{align*}
B_i(t,T):=\tfrac{1}{\alpha_i}(1-e^{-\alpha_i(T-t)}).
\end{align*}
\end{thm}
\begin{proof}
As in the proof of Theorem \ref{symmetric G-martingale under the equivalent sublinear expectation}, the assertion follows if we show that the process $X=(X_t)_{0\leq t\leq T}$, defined by
\begin{align*}
X_t:=\exp\Big(-\int_0^T\mu(s)ds-\sum_{i=1}^dB_i(t,T)X_t^i-\tfrac{1}{2}\sum_{i,j=1}^dB_i(t,T)B_j(t,T)q_t^{ij}-\sum_{i=1}^d\int_0^tX_s^ids\Big),
\end{align*}
is a symmetric $G$-martingale. Applying It\^o's formula for $G$-Brownian motion to $X$ yields
\begin{align*}
X_t=X_0+\int_0^t\Delta_uX_udu-\sum_{i=1}^d\int_0^tB_i(u,T)X_udB_u^i+\sum_{i,j=1}^d\int_0^t\Delta_u^{ij}X_ud\langle B^i,B^j\rangle_u,
\end{align*}
where the drift terms $\Delta=(\Delta_t)_{0\leq t\leq T}$ and $\Delta^{ij}=(\Delta_t^{ij})_{0\leq t\leq T}$, for all $i,j$, are given by
\begin{align*}
\Delta_t:={}&-\sum_{i=1}^d\partial_tB_i(t,T)X_t^i-\tfrac{1}{2}\sum_{i,j=1}^d\big(\partial_tB_i(t,T)B_j(t,T)+B_i(t,T)\partial_tB_j(t,T)\big)q_t^{ij}
\\&{}-\sum_{i=1}^dB_i(t,T)(-\alpha_iX_t^i)-\tfrac{1}{2}\sum_{i,j=1}^dB_i(t,T)B_j(t,T)\big(-(\alpha_i+\alpha_j)q_t^{ij}\big)-\sum_{i=1}^dX_t^i
\\={}&-\sum_{i=1}^d\big(\partial_tB_i(t,T)-\alpha_iB_i(t,T)+1\big)X_t^i
\\&{}-\tfrac{1}{2}\sum_{i,j=1}^dB_j(t,T)\big(\partial_tB_i(t,T)-\alpha_iB_i(t,T)\big)q_t^{ij}
\\&{}-\tfrac{1}{2}\sum_{i,j=1}^dB_i(t,T)\big(\partial_tB_j(t,T)-\alpha_jB_j(t,T)\big)q_t^{ij},
\\\Delta_t^{ij}:={}&-\tfrac{1}{2}B_i(t,T)B_j(t,T)+\tfrac{1}{2}B_i(t,T)B_j(t,T)=0,
\end{align*}
respectively. Since the function $B_i$, for all $i$, satisfies
\begin{align*}
\partial_tB_i(t,T)=\alpha_iB_i(t,T)-1
\end{align*}
and $q^{ij}=q^{ji}$ for all $i,j$, we obtain
\begin{align*}
X_t=X_0+\sum_{i=1}^d\int_0^tB_i(u,T)X_uq_u^idu-\sum_{i=1}^d\int_0^tB_i(u,T)X_udB_u^i.
\end{align*}
We can use the same argument as in the proof of Theorem \ref{symmetric G-martingale under the equivalent sublinear expectation} to show that $X_t\in L_G^2(\Omega_t)$ for all $t$. The Girsanov transformation for $G$-Brownian motion implies that the process $\bar{B}=(\bar{B}_t^1,...,\bar{B}_t^d)_{0\leq t\leq\tau}$, defined by
\begin{align*}
\bar{B}_t^i:=B_t^i-\int_0^tq_u^idu,
\end{align*}
is a $G$-Brownian motion under $\bar{\mathbb{E}}$. Therefore, the process $X$ is a symmetric $G$-martingale under $\bar{\mathbb{E}}$.
\end{proof}
\par We can use the function $\mu$ to fit the model to an initially observed term structure. Let us assume that there is a sufficiently regular forward curve $f_0^*:[0,\tau]\rightarrow\mathbb{R}$, which is currently observed on the market. Then one can check that the theoretical forward curve implied by the model matches the observed one at inception if and only if $\mu(t)=f_0^*(t)$ for all $t$.

\section{Conclusion}\label{conclusion}
In the present paper, we investigate the traditional Hull-White model when there is Knightian uncertainty about the volatility. We show that the common approach to pricing zero-coupon bonds, martingale modeling, does not work in the presence of volatility uncertainty. Hence, we follow a different approach. By introducing a market price of uncertainty, we adjust the short rate by its uncertain variance to obtain an arbitrage-free term structure. The resulting term structure is completely robust with respect to the volatility. The bond prices do neither depend on the future evolution of the volatility nor on its bounds. Instead, they depend on the current value of the market price of uncertainty. In particular, the bonds are exponentially affine with respect to the short rate and the market price of uncertainty. Due to the adjustment of the short rate, the model is inconsistent with the traditional Hull-White model. However, the model becomes consistent with the traditional one after fitting the model prices to the yield curve. All of these results hold true if the short rate is driven by multiple risk factors.
\par A first question for further research is whether the proposed term structure and the particular choice of the market price of uncertainty can be supported by an equilibrium in a representative agent economy. The present approach is purely based on no-arbitrage pricing. Instead, one could investigate a structural model in the spirit of \citet*{coxingersollross1985} under model uncertainty. \citet*{gagliardiniporchiatrojani2009} examine a structural model with ambiguity. The representative agent in the model faces ambiguity about the drift of the underlying risk factors. Since she is ambiguity averse, the agent has to solve a max-min expected utility problem. The solution determines the uncertain drift process, which is termed \textit{market price of ambiguity}. In a similar fashion, one could examine a structural model in which the representative agent faces ambiguity about the volatility. For this purpose one could adapt the framework of \citet*{epsteinji2013} to find out if there is an equilibrium in a representative agent economy supporting the specific market price of uncertainty used in the present model.
\par Apart from that, it would be interesting to test the empirical performance of the model, especially in comparison to traditional term structure models. One could, e.g., test if the model is able to explain the violation of the expectations hypothesis as it is done, for instance, by \citet*{daisingleton2003} and \citet*{gagliardiniporchiatrojani2009} and if it does better than traditional models. \citet*{famabliss1987}, \citet*{campbellshiller1991}, and \citet*{cochranepiazzesi2005} empirically test the expectations hypothesis by regressing changes in the yield curve onto the slope of the yield curve, which shows that the expectations hypothesis is violated. The regression produces negative coefficients, decreasing with respect to the maturity. \citet*{daisingleton2003} test the ability of several term structure models to explain the empirical findings. For this purpose, they fit the models to data and simulate term structures from the fitted models. Then they run regressions as above and compare the regression coefficients from the simulated data with the ones from real data. A successful model is supposed to match the coefficients from real data. In order to test the performance of the present model in this regard, one has to use a robust approach to calibrate the model as it is described at the end of Section \ref{yield curve fitting}. Moreover, one has to develope a simulation procedure which works in the presence of volatility uncertainty. The volatility is uncertain in the sense of Knightian uncertainty. By its definition, Knightian uncertainty cannot be measured by any probability. Thus, standard simulation procedures cannot be used. Instead, one has to construct a robust simulation procedure.

\bibliography{C:/Users/jhoelzermann/Documents/Uni/Literature/Literature}

\begin{thebibliography}{}

\bibitem[\protect\citeauthoryear{Acciaio, Beiglb{\"o}ck, Penkner, and
  Schachermayer}{Acciaio
  et~al.}{2016}]{acciaiobeiglbockpenknerschachermayer2016}
Acciaio, B., M.~Beiglb{\"o}ck, F.~Penkner, and W.~Schachermayer (2016).
\newblock A model-free version of the fundamental theorem of asset pricing and
  the super-replication theorem.
\newblock {\em Mathematical Finance\/}~{\em 26\/}(2), 233--251.

\bibitem[\protect\citeauthoryear{Adrian, Crump, and Moench}{Adrian
  et~al.}{2013}]{adriancrumpmoench2013}
Adrian, T., R.~K. Crump, and E.~Moench (2013).
\newblock Pricing the term structure with linear regressions.
\newblock {\em Journal of Financial Economics\/}~{\em 110\/}(1), 110--138.

\bibitem[\protect\citeauthoryear{Aksamit, Deng, Ob{\l}{\'o}j, and Tan}{Aksamit
  et~al.}{2019}]{aksamitdengoblojtan2019}
Aksamit, A., S.~Deng, J.~Ob{\l}{\'o}j, and X.~Tan (2019).
\newblock The robust pricing–hedging duality for {A}merican options in
  discrete time financial markets.
\newblock {\em Mathematical Finance\/}~{\em 29\/}(3), 861--897.

\bibitem[\protect\citeauthoryear{Avellaneda, Levy, and Par{\'a}s}{Avellaneda
  et~al.}{1995}]{avellanedalevyparas1995}
Avellaneda, M., A.~Levy, and A.~Par{\'a}s (1995).
\newblock Pricing and hedging derivative securities in markets with uncertain
  volatilities.
\newblock {\em Applied Mathematical Finance\/}~{\em 2\/}(2), 73--88.

\bibitem[\protect\citeauthoryear{Avellaneda and Lewicki}{Avellaneda and
  Lewicki}{1996}]{avellanedalewicki1996}
Avellaneda, M. and P.~Lewicki (1996).
\newblock Pricing interest rate contingent claims in markets with uncertain
  volatilities.
\newblock {\em Working Paper, Courant Institute of Mathematical Sciences\/}.

\bibitem[\protect\citeauthoryear{Bartl, Kupper, Pr{\"o}mel, and Tangpi}{Bartl
  et~al.}{2019}]{bartlkupperpromeltangpi2019}
Bartl, D., M.~Kupper, D.~J. Pr{\"o}mel, and L.~Tangpi (2019).
\newblock Duality for pathwise superhedging in continuous time.
\newblock {\em Finance and Stochastics\/}~{\em 23\/}(3), 697--728.

\bibitem[\protect\citeauthoryear{Bayraktar and Zhou}{Bayraktar and
  Zhou}{2017}]{bayraktarzhou2017}
Bayraktar, E. and Z.~Zhou (2017).
\newblock On arbitrage and duality under model uncertainty and portfolio
  constraints.
\newblock {\em Mathematical Finance\/}~{\em 27\/}(4), 988--1012.

\bibitem[\protect\citeauthoryear{Beiglb{\"o}ck, Cox, Huesmann, Perkowski, and
  Pr{\"o}mel}{Beiglb{\"o}ck
  et~al.}{2017}]{beiglbockcoxhuesmannperkowskipromel2017}
Beiglb{\"o}ck, M., A.~M.~G. Cox, M.~Huesmann, N.~Perkowski, and D.~J.
  Pr{\"o}mel (2017).
\newblock Pathwise superreplication via {V}ovk’s outer measure.
\newblock {\em Finance and Stochastics\/}~{\em 21\/}(4), 1141--1166.

\bibitem[\protect\citeauthoryear{Biagini, Bouchard, Kardaras, and Nutz}{Biagini
  et~al.}{2017}]{biaginibouchardkardarasnutz2017}
Biagini, S., B.~Bouchard, C.~Kardaras, and M.~Nutz (2017).
\newblock Robust fundamental theorem for continuous processes.
\newblock {\em Mathematical Finance\/}~{\em 27\/}(4), 963--987.

\bibitem[\protect\citeauthoryear{Bj{\"o}rk}{Bj{\"o}rk}{2004}]{bjork2004}
Bj{\"o}rk, T. (2004).
\newblock {\em Arbitrage Theory in Continuous Time}.
\newblock Oxford University Press.

\bibitem[\protect\citeauthoryear{Bj{\"o}rk, Di~Masi, Kabanov, and
  Runggaldier}{Bj{\"o}rk et~al.}{1997}]{bjorkdimasikabanovrunggaldier1997}
Bj{\"o}rk, T., G.~Di~Masi, Y.~Kabanov, and W.~Runggaldier (1997).
\newblock Towards a general theory of bond markets.
\newblock {\em Finance and Stochastics\/}~{\em 1\/}(2), 141--174.

\bibitem[\protect\citeauthoryear{Bouchard and Nutz}{Bouchard and
  Nutz}{2015}]{bouchardnutz2015}
Bouchard, B. and M.~Nutz (2015).
\newblock Arbitrage and duality in nondominated discrete-time models.
\newblock {\em The Annals of Applied Probability\/}~{\em 25\/}(2), 823--859.

\bibitem[\protect\citeauthoryear{Brigo and Mercurio}{Brigo and
  Mercurio}{2001}]{brigomercurio2001}
Brigo, D. and F.~Mercurio (2001).
\newblock {\em Interest Rate Models: Theory and Practice}.
\newblock Springer.

\bibitem[\protect\citeauthoryear{Burzoni, Frittelli, Hou, Maggis, and
  Ob{\l}{\'o}j}{Burzoni et~al.}{2019}]{burzonifrittellihoumaggisobloj2019}
Burzoni, M., M.~Frittelli, Z.~Hou, M.~Maggis, and J.~Ob{\l}{\'o}j (2019).
\newblock Pointwise arbitrage pricing theory in discrete time.
\newblock {\em Mathematics of Operations Research\/}~{\em 44\/}(3), 1034--1057.

\bibitem[\protect\citeauthoryear{Campbell and Shiller}{Campbell and
  Shiller}{1991}]{campbellshiller1991}
Campbell, J.~Y. and R.~J. Shiller (1991).
\newblock Yield spreads and interest rate movements: {A} bird's eye view.
\newblock {\em The Review of Economic Studies\/}~{\em 58\/}(3), 495--514.

\bibitem[\protect\citeauthoryear{Carassus, Ob{\l}{\'o}j, and Wiesel}{Carassus
  et~al.}{2019}]{carassusoblojwiesel2019}
Carassus, L., J.~Ob{\l}{\'o}j, and J.~Wiesel (2019).
\newblock The robust superreplication problem: {A} dynamic approach.
\newblock {\em SIAM Journal on Financial Mathematics\/}~{\em 10\/}(4),
  907--941.

\bibitem[\protect\citeauthoryear{Casassus, Collin-Dufresne, and
  Goldstein}{Casassus et~al.}{2005}]{casassuscollindufresnegoldstein2005}
Casassus, J., P.~Collin-Dufresne, and B.~Goldstein (2005).
\newblock Unspanned stochastic volatility and fixed income derivatives pricing.
\newblock {\em Journal of Banking \& Finance\/}~{\em 29\/}(11), 2723--2749.

\bibitem[\protect\citeauthoryear{Cochrane and Piazzesi}{Cochrane and
  Piazzesi}{2005}]{cochranepiazzesi2005}
Cochrane, J.~H. and M.~Piazzesi (2005).
\newblock Bond risk premia.
\newblock {\em The American Economic Review\/}~{\em 95\/}(1), 138--160.

\bibitem[\protect\citeauthoryear{Collin-Dufresne and Goldstein}{Collin-Dufresne
  and Goldstein}{2002}]{collindufresnegoldstein2002}
Collin-Dufresne, P. and R.~S. Goldstein (2002).
\newblock Do bonds span the fixed income markets? {T}heory and evidence for
  unspanned stochastic volatility.
\newblock {\em The Journal of Finance\/}~{\em 57\/}(4), 1685--1730.

\bibitem[\protect\citeauthoryear{Cont and Perkowski}{Cont and
  Perkowski}{2019}]{contperkowski2019}
Cont, R. and N.~Perkowski (2019).
\newblock Pathwise integration and change of variable formulas for continuous
  paths with arbitrary regularity.
\newblock {\em Transactions of the American Mathematical Society, Series
  B\/}~{\em 6}, 161--186.

\bibitem[\protect\citeauthoryear{Cox, Ingersoll~Jr., and Ross}{Cox
  et~al.}{1985}]{coxingersollross1985}
Cox, J.~C., J.~E. Ingersoll~Jr., and S.~A. Ross (1985).
\newblock A theory of the term structure of interest rates.
\newblock {\em Econometrica\/}~{\em 53\/}(2), 385--408.

\bibitem[\protect\citeauthoryear{Dai and Singleton}{Dai and
  Singleton}{2003}]{daisingleton2003}
Dai, Q. and K.~Singleton (2003).
\newblock Term structure dynamics in theory and reality.
\newblock {\em The Review of Financial Studies\/}~{\em 16\/}(3), 631--678.

\bibitem[\protect\citeauthoryear{Dai, Singleton, and Yang}{Dai
  et~al.}{2007}]{daisingletonyang2007}
Dai, Q., K.~J. Singleton, and W.~Yang (2007).
\newblock Regime shifts in a dynamic term structure model of {U.S.} {T}reasury
  bond yields.
\newblock {\em The Review of Financial Studies\/}~{\em 20\/}(5), 1669--1706.

\bibitem[\protect\citeauthoryear{Denis, Hu, and Peng}{Denis
  et~al.}{2011}]{denishupeng2011}
Denis, L., M.~Hu, and S.~Peng (2011).
\newblock Function spaces and capacity related to a sublinear expectation:
  {A}pplication to ${G}$-{B}rownian motion paths.
\newblock {\em Potential Analysis\/}~{\em 34\/}(2), 139--161.

\bibitem[\protect\citeauthoryear{Denis and Martini}{Denis and
  Martini}{2006}]{denismartini2006}
Denis, L. and C.~Martini (2006).
\newblock A theoretical framework for the pricing of contingent claims in the
  presence of model uncertainty.
\newblock {\em The Annals of Applied Probability\/}~{\em 16\/}(2), 827--852.

\bibitem[\protect\citeauthoryear{Epstein and Wilmott}{Epstein and
  Wilmott}{1999}]{epsteinwilmott1999}
Epstein, D. and P.~Wilmott (1999).
\newblock A nonlinear non-probabilistic spot interest rate model.
\newblock {\em Philosophical Transactions of the Royal Society of London A:
  Mathematical, Physical and Engineering Sciences\/}~{\em 357\/}(1758),
  2109--2117.

\bibitem[\protect\citeauthoryear{Epstein and Ji}{Epstein and
  Ji}{2013}]{epsteinji2013}
Epstein, L.~G. and S.~Ji (2013).
\newblock Ambiguous volatility and asset pricing in continuous time.
\newblock {\em The Review of Financial Studies\/}~{\em 26\/}(7), 1740--1786.

\bibitem[\protect\citeauthoryear{Fadina, Neufeld, and Schmidt}{Fadina
  et~al.}{2019}]{fadinaneufeldschmidt2019}
Fadina, T., A.~Neufeld, and T.~Schmidt (2019).
\newblock Affine processes under parameter uncertainty.
\newblock {\em Probability, Uncertainty and Quantitative Risk\/}~{\em 4\/}(5).

\bibitem[\protect\citeauthoryear{Fadina and Schmidt}{Fadina and
  Schmidt}{2019}]{fadinaschmidt2019}
Fadina, T. and T.~Schmidt (2019).
\newblock Default ambiguity.
\newblock {\em Risks\/}~{\em 7\/}(2), 64.

\bibitem[\protect\citeauthoryear{Fama and Bliss}{Fama and
  Bliss}{1987}]{famabliss1987}
Fama, E.~F. and R.~R. Bliss (1987).
\newblock The information in long-maturity forward rates.
\newblock {\em The American Economic Review\/}~{\em 77\/}(4), 680--692.

\bibitem[\protect\citeauthoryear{Fong and Vasicek}{Fong and
  Vasicek}{1991}]{fongvasicek1991}
Fong, H.~G. and O.~A. Vasicek (1991).
\newblock Fixed-income volatility management.
\newblock {\em The Journal of Portfolio Management\/}~{\em 17\/}(4), 41--46.

\bibitem[\protect\citeauthoryear{Gagliardini, Porchia, and Trojani}{Gagliardini
  et~al.}{2009}]{gagliardiniporchiatrojani2009}
Gagliardini, P., P.~Porchia, and F.~Trojani (2009).
\newblock Ambiguity aversion and the term structure of interest rates.
\newblock {\em The Review of Financial Studies\/}~{\em 22\/}(10), 4157--4188.

\bibitem[\protect\citeauthoryear{Gourieroux, Monfort, Pegoraro, and
  Renne}{Gourieroux et~al.}{2014}]{gourierouxmonfortpegorarorenne2014}
Gourieroux, C., A.~Monfort, F.~Pegoraro, and J.-P. Renne (2014).
\newblock Regime switching and bond pricing.
\newblock {\em Journal of Financial Econometrics\/}~{\em 12\/}(2), 237--277.

\bibitem[\protect\citeauthoryear{Hu, Ji, Peng, and Song}{Hu
  et~al.}{2014}]{hujipengsong2014}
Hu, M., S.~Ji, S.~Peng, and Y.~Song (2014).
\newblock Comparison theorem, {F}eynman--{K}ac formula and {G}irsanov
  transformation for {BSDE}s driven by ${G}$-{B}rownian motion.
\newblock {\em Stochastic Processes and their Applications\/}~{\em 124\/}(2),
  1170--1195.

\bibitem[\protect\citeauthoryear{Joslin, Priebsch, and Singleton}{Joslin
  et~al.}{2014}]{joslinpriebschsingleton2014}
Joslin, S., M.~Priebsch, and K.~Singleton (2014).
\newblock Risk premiums in dynamic term structure models with unspanned macro
  risks.
\newblock {\em The Journal of Finance\/}~{\em 69\/}(3), 1197--1233.

\bibitem[\protect\citeauthoryear{Klein, Schmidt, and Teichmann}{Klein
  et~al.}{2016}]{kleinschmidtteichmann2016}
Klein, I., T.~Schmidt, and J.~Teichmann (2016).
\newblock No arbitrage theory for bond markets.
\newblock In {\em Advanced Modelling in Mathematical Finance}, pp.\  381--421.
  Springer.

\bibitem[\protect\citeauthoryear{Knight}{Knight}{1921}]{knight1921}
Knight, F.~H. (1921).
\newblock {\em Risk, Uncertainty, and Profit}.
\newblock Hart, Schaffner \& Marx, New York.

\bibitem[\protect\citeauthoryear{Li and Peng}{Li and Peng}{2011}]{lipeng2011}
Li, X. and S.~Peng (2011).
\newblock Stopping times and related {I}t{\^o}’s calculus with
  ${G}$-{B}rownian motion.
\newblock {\em Stochastic Processes and their Applications\/}~{\em 121\/}(7),
  1492--1508.

\bibitem[\protect\citeauthoryear{Longstaff and Schwartz}{Longstaff and
  Schwartz}{1992}]{longstaffschwartz1992}
Longstaff, F.~A. and E.~S. Schwartz (1992).
\newblock Interest rate volatility and the term structure: {A} two-factor
  general equilibrium model.
\newblock {\em The Journal of Finance\/}~{\em 47\/}(4), 1259--1282.

\bibitem[\protect\citeauthoryear{Lyons}{Lyons}{1995}]{lyons1995}
Lyons, T.~J. (1995).
\newblock Uncertain volatility and the risk-free synthesis of derivatives.
\newblock {\em Applied Mathematical Finance\/}~{\em 2\/}(2), 117--133.

\bibitem[\protect\citeauthoryear{Monfort and Pegoraro}{Monfort and
  Pegoraro}{2007}]{monfortpegoraro2007}
Monfort, A. and F.~Pegoraro (2007).
\newblock Switching {VARMA} term structure models.
\newblock {\em Journal of Financial Econometrics\/}~{\em 5\/}(1), 105--153.

\bibitem[\protect\citeauthoryear{Nutz}{Nutz}{2013}]{nutz2013}
Nutz, M. (2013).
\newblock Random ${G}$-expectations.
\newblock {\em The Annals of Applied Probability\/}~{\em 23\/}(5), 1755--1777.

\bibitem[\protect\citeauthoryear{Peng}{Peng}{2019}]{peng2019}
Peng, S. (2019).
\newblock {\em Nonlinear Expectations and Stochastic Calculus under
  Uncertainty}.
\newblock Springer.

\bibitem[\protect\citeauthoryear{Possama{\"\i}, Royer, and Touzi}{Possama{\"\i}
  et~al.}{2013}]{possamairoyertouzi2013}
Possama{\"\i}, D., G.~Royer, and N.~Touzi (2013).
\newblock On the robust superhedging of measurable claims.
\newblock {\em Electronic Communications in Probability\/}~{\em 18\/}(95).

\bibitem[\protect\citeauthoryear{Riedel}{Riedel}{2015}]{riedel2015}
Riedel, F. (2015).
\newblock Financial economics without probabilistic prior assumptions.
\newblock {\em Decisions in Economics and Finance\/}~{\em 38\/}(1), 75--91.

\bibitem[\protect\citeauthoryear{Schied and Voloshchenko}{Schied and
  Voloshchenko}{2016}]{schiedvoloshchenko2016}
Schied, A. and I.~Voloshchenko (2016).
\newblock Pathwise no-arbitrage in a class of delta hedging strategies.
\newblock {\em Probability, Uncertainty and Quantitative Risk\/}~{\em 1\/}(3).

\bibitem[\protect\citeauthoryear{Soner, Touzi, and Zhang}{Soner
  et~al.}{2011}]{sonertouzizhang2011b}
Soner, H.~M., N.~Touzi, and J.~Zhang (2011).
\newblock Quasi-sure stochastic analysis through aggregation.
\newblock {\em Electronic Journal of Probability\/}~{\em 16\/}(67), 1844--1879.

\bibitem[\protect\citeauthoryear{Song}{Song}{2011}]{song2011}
Song, Y. (2011).
\newblock Some properties on {G}-evaluation and its applications to
  {G}-martingale decomposition.
\newblock {\em Science China Mathematics\/}~{\em 54\/}(2), 287--300.

\bibitem[\protect\citeauthoryear{Vorbrink}{Vorbrink}{2014}]{vorbrink2014}
Vorbrink, J. (2014).
\newblock Financial markets with volatility uncertainty.
\newblock {\em Journal of Mathematical Economics\/}~{\em 53}, 64--78.

\end{thebibliography}
\bibliographystyle{chicago}

\end{document}